\documentclass[aps,prb,twocolumn,showpacs,10pt]{revtex4-1} 
\usepackage{amssymb}
\usepackage[pdftex]{graphicx}
\usepackage{amsmath,amsfonts,amssymb,dsfont,color,bbm}
\usepackage{tikz}

\usepackage[colorlinks]{hyperref}

\newcommand{\ie}{\emph{i.e.}}

\newcommand{\avg}[1]{\overline{#1}}

\newcommand{\be}{\begin{equation}}
\newcommand{\ee}{\end{equation}}
\newcommand{\bsigma}{\boldsymbol\sigma}
\newcommand{\bk}{\boldsymbol{k}}
\newcommand{\bp}{\boldsymbol{p}}

\newcommand{\eps}{\varepsilon}

\newcommand{\R}{\mathbbm{R}}
\newcommand{\tr}{\mathrm{tr}}
\renewcommand{\Re}{\mathrm{Re}}
\newcommand{\Z}{\mathbbm{Z}}

\newcommand{\tmatrix}[2]{\begin{bmatrix} #1 \\ #2 \end{bmatrix}}

\begin{document}
\title{Localization behavior 
of Dirac particles in disordered graphene superlattices}

\author{Qifang Zhao}
\affiliation{Department of Physics and Centre for Computational Science and Engineering, National University of Singapore, Singapore 117546, Republic of Singapore}
\author{Jiangbin Gong} \email{phygj@nus.edu.sg}
\affiliation{Department of Physics and Centre for Computational Science and Engineering, National University of Singapore, Singapore 117546, Republic of Singapore}
\author{Cord A.\ M\"uller} 
\affiliation{Centre for Quantum Technologies, National University of
Singapore, Singapore 117543, Republic of Singapore}


\begin{abstract}
Graphene superlattices (GSLs), formed by subjecting a monolayer graphene sheet
to a periodic potential, can be used to engineer band structures
and, from there, charge transport properties, but these are sensitive
to the presence of disorder.  
The localization behavior of massless 2D Dirac particles induced by weak disorder is
studied  for both scalar-potential and vector-potential GSLs, computationally as well as analytically by a weak-disorder
expansion. In particular, it is investigated how the Lyapunov exponent (inverse localization length)
depends on the incidence angle to a 1D GSL. 
Delocalization resonances are found for both scalar and vector GSLs.
The sharp angular dependence of the Lyapunov exponent may be exploited
to realize disorder-induced filtering, as verified by full 2D
numerical wave packet simulations.  
\end{abstract}


\maketitle
\section{Introduction}

One fundamental  aspect of graphene lies in the linear dispersion relation
of its low-energy charge carriers (electrons and holes) around the so-called Dirac points.
These charge carriers behave as relativistic massless chiral Dirac fermions and can be described
by a two-dimensional (2D) Dirac equation. 
\cite{Weiss1958, Semenoff1984, Haldane1988} 
The linear dispersion relation
is responsible for many discoveries in recent graphene
research,\cite{Geim2009} such as 
half-integer quantum Hall effect,\cite{Sharapov2005,Kim2006}
Klein's paradox,\cite{Klein1929,Katsnelson2006}
and Zitterbewegung. \cite{Zitter} 
Other than to graphene, Dirac or Dirac-like equations
naturally apply to cold atoms, \cite{Ohberg2008,Lee2009,Oh2010,Pachos2011,Alexandrov2011}  trapped
ions, \cite{Lamata2011} 
semiconductors, \cite{Zawadzki2011} or polaritons. \cite{Unanyan2010}

Motivated by the importance of Dirac equations in such a wide variety of frontier research areas, we study in this work
disorder-induced localization~\cite{Anderson1958,Kramer1993} of
massless Dirac particles
in random potentials.  Though our results are presented in the
context of disordered 
graphene superlattices (GSLs, see below) we expect them to be
useful for many other settings as well.
For example, when disorder is introduced to cold-atom simulations
of graphene \cite{Lee2009} or GSLs, \cite{Alexandrov2011} our general
treatment can be adapted to study the impact of randomness on
the transport of Dirac matter waves.

GSL refers to graphene under external periodic scalar\cite{X_Zhang2007,Louie2008,Pereira2008,Louie2008_2,Lin2009,
Fertig2009,Peeters2010,SYZhu2010,Louie2011} or vector potentials.\cite{Louie2011,Peeters2008,
Peeters2009,Sharma2009,Martino2009,SJYang2008,Snyman2009,Louie2010,Titov2010}
Because GSLs further tailor the band dispersion relation of graphene,
they may be used to construct graphene-based quantum devices.
Theoretical studies of GSLs and graphene under periodic
corrugation\cite{Jonson2008,Vozmediano2008,Lichtenstein2008} have been
highly fruitful, with remarkable 
findings such as electron beam supercollimation \cite{Louie2008_2} and the
emergence of extra Dirac
points. \cite{Fertig2009,Peeters2010,SYZhu2010,Louie2011}  
On the experimental side, GSLs with scalar potential barriers
can be created via the electric field effect or chemical doping. \cite{Nov2004,Firsov2005,Kim2005}
Two-dimensional (2D) GSLs with a period as small as  $5\,$nm have been created through electron-beam induced
deposition of carbon. \cite{Zettl2008} 
Also triangular GSLs growing on different metal surfaces
have been observed.
\cite{Wintterlin2007,Miranda2008,Sutter2008,Greber2008,Michely2008,Michely2008_2,Michely2009}
Besides, nano-ripple arrays are generated by chemical vapor deposition (CVD). \cite{Avsar2011} 
Vector potentials are induced by magnetic
fields\cite{Egger2007} or physical strain, \cite{Neto2009}
so vector GSLs can be realized by mounting
graphene on a substrate with a periodic array of ferromagnetic strips or
a periodically structured substrate. \cite{Peeters2009_3}

All these laboratory-produced GSLs cannot be perfectly periodic, due to
intrinsic randomness and uncontrollable factors during production.
Therefore, a more realistic GSL should be modeled by a periodic
potential plus some weak disorder in potential height, potential
width, or lattice spacing. This randomness causes Anderson
localization,  which
turns conductors into insulators and is especially severe in low dimensions. \cite{Anderson1958,Kramer1993} 
Consequently, the focus of our work is on
the localization behavior of a 2D Dirac particle in
weakly disordered 1D GSLs.
In a related work, \cite{ZDWang2009}  localization 
of Dirac particles in 1D disordered potentials was studied, but
only for zero incidence angle $\theta$ (i.e., wave vector of charge
carriers normal to the interface between different GSL layers) and without analytical results for the localization length. 
Another closely related theoretical study of disordered GSLs 
\cite{Nori2009} comprised an analytical discussion of the scattering transmission only 
for sufficiently small $\theta$ 
and random barrier heights. Our work extends all previous results, to the best of our knowledge,
inasmuch as it covers the analytical properties of the localization
length for all values of $\theta$, for different types of disorder,
and for both scalar and vector GSLs.  

The paper is organized as follows. In Sec.~II,  we begin by modeling disordered scalar and vector
GSLs by 1D rectangular potential barriers or wells. Using
a transfer matrix formalism, 
we then derive the weak-disorder expansion of the localization length,
or equivalently the associated Lyapunov exponent. 
In Sec.~III we present analytical and numerical results
for the Lyapunov exponent of scalar GSLs, as modeled by disordered delta or
rectangular potentials. It is found that at fixed energy, 
the localization length depends very intricately upon
the incidence angle $\theta$ of 2D Dirac particles in the graphene
plane.  
We also predict and confirm the existence
of delocalization resonances other than for perpendicular incidence: along these directions the
Lyapunov exponent vanishes.  Our theoretical predictions are fully supported by numerical
results, as also reported below.  
Section IV is in parallel with Sec.~III, but treats
GSLs with vector potentials.
In addition, assisted by a numerical
study of wave-packet dynamics in Sec.~V, we propose to
use the angular dependence of the localization length to realize a
disorder-based filtering mechanism.  
 Section VI
concludes.


\section{Localization length in disordered graphene superlattices}


\subsection{Disordered graphene superlattices}

Thanks to their linear dispersion relation, low-energy charge carriers
near the Dirac points in graphene are well described by the
2D massless Dirac Hamiltonian:
\be
    \label{First}
    H =  v_F \bsigma\cdot\bp + V(x).
\ee
In graphene, $v_F\approx 10^{6}\,$m/s is the Fermi velocity;
$\bsigma\equiv (\sigma_{x},\sigma_{y})$ is the vector of Pauli
matrices.
We consider a graphene superlattice (GSL) of parallel potential
barriers, such that the external potential $V(x)$ depends only on
$x$.
In the following, we consider both scalar and vector superlattices.

\begin{figure}
   \centering
\includegraphics[width=\linewidth]{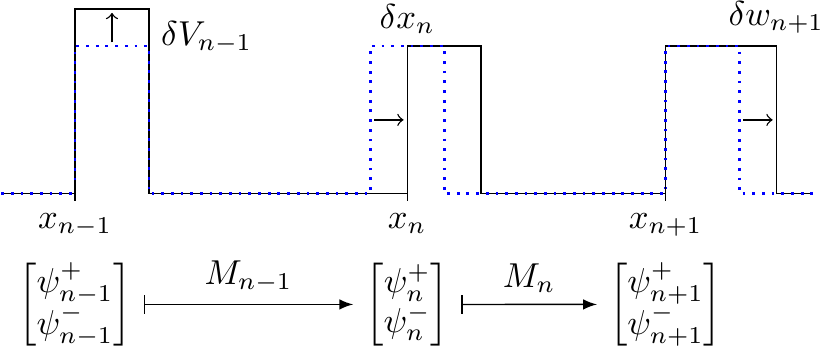}
\caption{Top: Disordered graphene superlattice (GSL)
     realized as a scalar potential, Eq.~\eqref{scalarpot}, or vector potential,
      Eq.~\eqref{vectorpot}.
 Deviations from the clean GSL (dotted) can occur via fluctuations in barrier height $\delta
     V$, barrier width $\delta w$, lattice spacing $\delta l$, and
   combinations thereof. Bottom: Right- and left-travelling
   wavefunction amplitudes are mapped from one barrier to the other by
   the transfer matrix, Eq.~\eqref{Mn}.
}
   \label{fig:KP}
\end{figure}

A general scalar
superlattice potential can be described by
\be\label{scalarpot}
V(x) = \sum_{n\in\Z} V_n(x-x_n).
\ee
We will consider rectangular potential barriers (or wells) as depicted
in Fig.~\ref{fig:KP}. A perfect GSL has identical potential barriers (or wells) of height
$V$ and width $w$, i.e.\
$V_n(x) = V$ if $0<x<w$ and 0 elsewhere, at lattice positions $x_n= n l$.  Due to unavoidable experimental
imperfections, or deliberate introduction of randomness, these
potential parameters 
fluctuate from site to site:
\begin{align}
V_n &= V+\delta V_n ,\\
x_n & = nl +\delta x_n, \\
w_n &= w+\delta w_n.
\end{align}
This randomness can induce localization, as will be discussed at
length in Sec.~\ref{scalarpot.sec}. 

A vector-potential superlattice is defined in terms of the
matrix-valued potential
\be \label{vectorpot}
    V(x)= -\sigma_y  \frac{e v_F}{c} \sum_{n\in\Z}{A_n (x-x_n)}.
\ee
Defining $V_n =  e v_F A_n/c$ and assuming $A_n(x-x_n)$ is of the same
form as $V_n(x-x_n)$, one deals with the same parameters
as in the scalar case. The different
potential nature, however, implies very different
localization properties, as will become clear in Sec.~\ref{vectorpot.sec}.

\subsection{Transfer-matrix formalism}

Because the potential $V(x)$ is separable, the problem of describing
the transmission across the lattice is effectively 1D,
and the transfer matrix formalism is
particularly suited.\cite{Soukoulis2008}

The scattering of a massless Dirac particle through a single square
barrier (well) is well understood,  for scalar as well as vector
potentials. \cite{Katsnelson2006,Neto2009}
Since the potential is piecewise constant, the solution to the Dirac
equation is a plane wave, both inside and outside the barrier.
Outside the barrier, solutions of energy $E=s\hbar v_F k$
with $s=\pm 1$ and $k=|\bk| = (k_x^2+k_y^2)^{1/2}$  are the Dirac
bispinors
\begin{equation} \label{freePsi}
\Psi^{\pm}(x,y) = e^{\pm i k_x  x  + i k_y y}  \begin{pmatrix} 1 \\  \pm
s e^{\pm i\theta} \end{pmatrix},
\end{equation}
travelling towards
right ($+$) and left ($-$), with $k_x\ge0$ by convention.
\be
\theta=\tan^{-1} \frac{k_y}{k_x}
\ee
is the incidence angle, or angle of propagation (outside the barrier) with respect to the $x$-axis.

In the lattice,  the wave function
between barriers, where $V(x)=0$, is a superposition of free right- and
left-moving components created by repeated elastic reflexion and transmission.
It is useful to parametrize the wave function on the left side of the $n$th barrier, $\Psi_n =
\lim_{\epsilon\to0^+}\Psi(x_n-\epsilon)$, as
\be
\Psi_n = \psi^+_n \begin{pmatrix} 1 \\ s e^{i
\theta}\end{pmatrix} e^{ik_y y} + \psi^-_n \begin{pmatrix} 1 \\
-s e^{-i \theta}\end{pmatrix} e^{ik_y y}.
\ee
Since the free solutions \eqref{freePsi} between barriers
are fixed, scattering cannot mix the two components of the
bispinor, and it suffices to introduce the two amplitudes
$\psi^{\pm}_n$, just as for a 
scalar wave obeying Schr\"odinger's
equation on a 1D lattice.  
These amplitudes are mapped from $n$ to $n+1$ by the transfer matrix:
\be
    \tmatrix{\psi_{n+1}^+}{\psi_{n+1}^-}
 = M_n
    \tmatrix{\psi_{n}^+}{\psi_{n}^-}
\ee
with
\be
\label{Mn}
M_n = \begin{bmatrix}
             \dfrac{1}{t_n^*} e^{i\Delta_n}& - \dfrac{r_n^*}{t_n^*}e^{i\Delta_n} \\
            -\dfrac{r_n}{t_n} e^{-i\Delta_n}& \dfrac{1}{t_n}e^{-i\Delta_n}
      \end{bmatrix}.
\ee
Reflection and transmission amplitudes $r_n$ and $t_n$ are known
functions of barrier parameters $\{V_n,x_n,w_n\}$ and quantum numbers $\{k_x,k_y,s\}$ or equivalently
$\{E,\theta,s\}$\cite{Katsnelson2006,Neto2009}. 
$\Delta_n \equiv k_x (x_{n+1}-x_n)$ is the free propagation phase
between superlattice points in the absence of any barriers.  
The transfer matrix is largely determined by
the symmetries of the scattering
problem. \cite{Soukoulis2008,Delande2010} 
Unitarity or current conservation
implies $\det M_n = 1 = |r_n|^2 +
|t_n|^2$. Thus the total reflection and
transmission probabilities can be expressed as $R_n= |r_n|^2 = \sin^2\phi_n$ and $T_n=
|t_n|^2 = \cos^2\phi_n$, and we find it useful to
parameterize $M_n$ as
\be
  \label{Para}
   M_n        = \begin{bmatrix} e^{i\alpha_n} \sec \phi_n  & e^{i\beta_n} \tan \phi_n  \\
    e^{-i\beta_n} \tan \phi_n  & e^{-i\alpha_n} \sec \phi_n
  \end{bmatrix}.
\ee
By construction, the net transfer matrix across $N$ barriers is the product
\be
\label{PN}
P_N = \prod_{n=1}^N M_n.
\ee
Before studying this product for the random matrices $M_n$ arising from disorder, we first discuss its
implications for clean GSLs.

\subsection{Clean graphene superlattices}
In a clean GSL, all transfer matrices $M_n=M$ are identical. In other words, a single transfer matrix contains all
information about the dispersion relation in the lattice, which is the
essence of Bloch's theorem. 

If parameters are such that $|\tr M| < 2$, the energy $E$ lies within the conduction band of the
GSL.
In this case the eigenvalues of the transfer matrix are of the form $\lambda_\pm =
e^{\pm i\mu}$ with $\mu\in \R$, such that
$\tr M = 2\cos\mu$. The transfer phase $\mu = K_xl$ across one lattice
cell determines the Bloch vector $K_x$ of the extended solution in the
$x$ direction.
In terms of the parametrization Eq.~\eqref{Para} the
dispersion relation in the clean GSL
therefore reads \cite{Peeters2010,Soukoulis2008}
\be\label{Dispersion0}
    \cos K_x l = \cos \mu = \sec\phi \cos\alpha.
\ee
The structure of this dispersion is analogous to that of the
Kronig-Penny model,
\cite{GalindoPascualQM1} from which it differs
only in the functional dependence of
the transfer parameters $\{\phi,\alpha\}$ on the
potential parameters $\{V,l,w\}$ and $\{E,\theta,s\}$. This dependence will be made
explicit for the two cases of scalar and vector potentials in
Secs.~\ref{scalarpot.sec} and \ref{vectorpot.sec}, respectively.

In the case $|{\tr M}| > 2$, the energy $E$ falls into a
band gap. The wave cannot propagate, and $|{\tr M}| = 2 \cosh\kappa_x l$ defines the exponential decay rate
$\gamma = \kappa_x l$ across one
lattice cell. This characteristic
localization exponent
\be\label{gamma_gap}
\gamma = \ln |\lambda_+|
\ee
is determined by the larger one of the two eigenvalues of the
transfer matrix $M$, which also defines the Lyapunov exponent of
the product $P_N = M^N$, whose larger eigenvalue grows like
$\exp\{\gamma N\}$.


\subsection{Disordered graphene superlattices}
\label{cleanGSL.sec}

The transmission across a disordered lattice is described by the
product $P_N$ of random matrices shown in Eq.~\eqref{PN}.
Since the pioneering work of Furstenberg, \cite{Furstenberg} it is well
known that the larger eigenvalue of such a product grows
exponentially with probability one. This implies that a wave
incident on the disordered GSL at barrier number
$1$ has an
exponentially small probability of transmission after barrier number
$N$, which is one of the hallmarks of disorder-induced
localization. Indeed, at the first barrier, the wave splits into reflected and
transmitted components, and so on across the lattice. The boundary
condition is actually simpler after the last barrier $N$, where there is only the
transmitted component, but no component is incident from the right.
Starting with the reverse boundary condition (such as $\psi_{1}^+=1$
and $\psi_1^-=0$) at the left, the product in Eq.~\eqref{PN} predicts that the solution grows like
\be\label{PN11}
   |\psi_N^+|^2 = |(P_N)_{11}|^{2} \sim \exp\{2N \gamma\},
\ee
which suggests the expected exponential localization.
The Lyapunov exponent, mathematically defined as
\be
   \label{Lya}
   \gamma= \lim_{N\to\infty} \frac{1}{2N} \ln |(P_N)_{11}|^2 =
   -\lim_{N\to\infty} \frac{1}{2N} \ln T_N,
\ee
thus determines the localization length $l_\text{loc}
= l/\gamma$.
Here $T_N\in [0,1]$ is the net transmission probability after $N$
barriers.

The transmission is a random variable, with a very
wide probability distribution for long enough samples.  In the localized regime, its
most probable (or typical) value differs vastly from its mean. The extinction
$|{\ln T(N)}|$, however, has a probability distribution that converges
towards a normal distribution, such that its most probable value is
equal to the mean, and the right hand side of Eq.~\eqref{Lya} indeed
converges to the Lyapunov exponent. \cite{Delande2010,Abrikosov1981}

While it is an elementary exercise to multiply random matrices and
extract the Lyapunov exponent numerically, there is no simple, general method of calculating the Lyapunov
exponent exactly for a given model of disorder with arbitrary energy.
Different situations require different approaches.
In the following, we treat two different cases that are relevant in the GSL
context and allow for
analytical calculations.


\subsection{Randomly spaced, identical barriers}
\label{randomposition.sec}

First we consider the very simple case where identical barriers are
distributed with random positions such that
the free propagation phase 
between barriers  is
uniformly distributed in $[0,2\pi]$.  Under an ensemble average
$\avg{(\cdot)}$ over
these random phases, the extinction $|{\ln
    T_N}|$ across $N$ barriers is found to be additive along the
sample:
$\avg{\ln T_N} = N \ln
T_1$.\cite{Berry1997,Delande2010}
Here,
$T_1$ is the single-barrier transmission at
given energy $E$ and propagation angle $\theta$.
Equation~\eqref{Lya} then immediately yields the  Lyapunov exponent
  $  \gamma = -\frac{1}{2} \ln T_1$.
This result holds as long as the phases are random enough to satisfy the
assumption of a uniform distribution, 
but no matter how small $T_1$, 
i.e.\ how strong the scattering.

At a given energy, a rectangular barrier becomes perfectly transmitting
at certain incident angles, and notably at perpendicular incidence
($\theta=0$) for all energies---this phenomenon is known as Klein tunneling.\cite{Katsnelson2006}
In these cases, $T_1=1$ implies of course $\gamma=0$ and
absence of localization, because all barriers share the same resonance
condition.

\subsection{Weak-disorder expansion}

Although the previous elementary model captures the essence of disorder-induced
exponential localization, it cannot describe the more interesting, and
arguably more relevant, case of 
barriers with slightly random width, height, and/or spacing.  
In the following, we adapt the weak-disorder expansion of Derrida et
al.\cite{Derrida1987} to our case. Here, we describe briefly the steps
leading to the main result; details can be found in Appendices. 

First, we  Taylor-expand
\be
   \label{Mn2}
   M_n = M + \epsilon_n M' + \frac{\epsilon_n^2}{2}M'' + O(\epsilon_n^3),
\ee
where $M$ is the transfer matrix of the corresponding clean GSL, the prime $(.)'$
indicates differentiation with respect to the perturbed variable
$V$, $w$ or $d$), and $\epsilon_n$ is the weak perturbation ($\epsilon_n =
\delta V_n$, $\delta w_n$ or $\delta d_n$). We assume that the random variables at
different sites are independent and
identically distributed, with zero mean and finite variance:
\begin{align}
\avg{\epsilon} & = \lim_{N\to\infty}\frac{1}{N} \sum_{n=1}^N\epsilon_n
=0, \quad  \label{avgeps}\\
\avg{\epsilon^2} & =   \lim_{N\to\infty}\frac{1}{N}
\sum_{n=1}^N\epsilon_n ^2 \ge 0. \label{vareps}
\end{align}

Second, we expand
the product \eqref{PN} to order $\epsilon^2$
in the eigenbasis of $M$, where
$\tilde M 
= \text{diag}(\lambda_+,\lambda_-)$. The eigenbasis of $M$ can be used
to find the Lyapunov exponent because
the exponential growth rate is independent of the representation. The matrix element required in
\eqref{PN11} then reads, neglecting terms of order $\epsilon^3$,
\begin{align}
  \label{tildePN}
  (\tilde P_N)_{11} = \lambda_+^N \bigg[ 1
& +  \lambda_+^{-1}  \sum_{n =1}^N \left\{ \epsilon_n \tilde M'_{11}   +
\frac{\epsilon_n^2}{2} \tilde M''_{11} \right\}
 \\
 &  +   \sum_{n<m}
\epsilon_n\epsilon_m \lambda_+^{n-m-1}
(\tilde M'\tilde M^{m-n-1}\tilde M')_{11}
\bigg]. \nonumber
\end{align}
The second line involves only fluctuations at different sites and gives no contribution 
after the ensemble average (see App.~\ref{lyap1.apx} for details).
Inserting the first line 
into Eq.~\eqref{Lya}, and further using
Eqs.~\eqref{avgeps} and \eqref{vareps}, one obtains the
disorder-induced Lyapunov exponent
 \be
   \label{LyaM}
   \gamma = \frac{l}{l_\text{loc}} = \frac{\avg{\epsilon^2}}{2}
\Re\left\{ \lambda_+^{-1}\tilde M''_{11}  - \lambda_+^{-2}\left(\tilde M'_{11}\right)^2
\right\} .
\ee

In a third step we perform the 
diagonalization from $M$ to $\tilde M$ in order to arrive at an explicit expression as function of the
system variables (see App.~\ref{lyap2.apx} for details). In terms of
the parametrization \eqref{Para}
one obtains a relatively compact result:
\be
   \label{LyaF}
   \gamma
          = \frac{\avg{\epsilon^2}}{2}
\left\{
\frac{\tan^4\phi}{\sin^2\mu}\left[\left(\frac{\sin\alpha}{\sin\phi}\right)'\right]^2  + \beta'^2 \tan^2\phi\right\}.
\ee
Here, $\sin\mu$ is a function of
$\{\alpha,\phi\}$ via the clean dispersion relation
\eqref{Dispersion0}, which is assumed to be satisfied by a propagating
solution of energy $E$ (otherwise, this perturbative result of order
$\avg{\epsilon^2}$ is merely a small correction to the band-gap extinction of Sec.~\ref{cleanGSL.sec}).

Before we discuss the localization exponent 
\eqref{LyaF} in detail for scalar and vector potentials
(Secs.~\ref{scalarpot.sec} and \ref{vectorpot.sec}), we comment
on its limit of validity. 
Eq.~\eqref{LyaF} diverges at the band edges, where $\sin\mu=0$. It is
well known that localization at these special points occurs with an
anomalous localization length that differs from the perturbative
result.\cite{Derrida1984,Izrailev1998}
However, exponential localization
in the conduction band away from these special points is very well described by Eqs.~\eqref{LyaM}
  and \eqref{LyaF}, as will be checked via numerical calculations
  below.


\section{Scalar Potential}
\label{scalarpot.sec}

Now we specify the transfer-matrix parametrization in Eq.~\eqref{Para}
for a single scalar potential barrier, the building block of the scalar GSL, in order
to analyze the  Lyapunov exponent given in Eq.~\eqref{LyaF}.
The reflection and transmission amplitudes $r$ and $t$ are found
by piecing together a continuous plane-wave solution across the barrier:\cite{Katsnelson2006}
\begin{align}
\frac{1}{t} & = e^{iw k_x}
\left[
\cos\varphi +i s \sin\varphi  \frac{v -
  \eps \cos^2\theta}{ql \cos\theta }
\right],
\label{scalar1overtn}\\
\frac{r}{t} & = - s  e^{i
  w k_x} e^{i\theta} \tan\theta \frac{v\sin\varphi}{ql} .
\label{scalarrnovertn}
\end{align}
Here $\eps = E l/\hbar v_F = s|\bk|l $ and  $v = V l/\hbar v_F$ are energy and
barrier height expressed in lattice units. Furthermore,
$\varphi = q w $ is the phase picked
up by the plane wave with  wavevector $q =
[(v-\eps)^2l^{-2} - k_y^2]^{1/2}$ in the $x$-direction across the potential barrier.

In the next Sec.~\ref{deltascalarpot.sec}, we first discuss the limiting case of $\delta$-like barriers, which admits
simple expressions and helps to guide the understanding of the general
case, tackled in Sec.~\ref{generalscalarpot.sec}.

\subsection{Amplitude-disordered delta scalar potential}
\label{deltascalarpot.sec}

Consider an amplitude-disordered Dirac-Kronig-Penney model, made out of
regularly spaced $\delta$-peaks of random strength, or $\delta$GSL for
short. This description is appropriate in the low-energy regime, where
barriers become very narrow and high, $k_x w \ll 1 $ and
$v \gg \eps $.
 In the limit $w\to0$ and $v\to\infty$ at fixed $vw/l = \varphi$,
one has $ql\to v$, and the reflection and transmission coefficients
in Eqs.~\eqref{scalar1overtn} and \eqref{scalarrnovertn} become
\begin{align}
\frac{1}{t} & =\cos\varphi +i s \frac{\sin\varphi}{\cos\theta },
\label{scalar1overtn_delta}\\
\frac{r}{t} & = - s e^{i\theta} \tan\theta \sin\varphi.
\label{scalarrnovertn_delta}
\end{align}
These expressions depend on the barrier parameters only via the
combination $\varphi = vw/l$. Therefore, they cover randomness in both barrier width and
height. We assume a resulting phase-shift distribution with mean
$\varphi=\avg{\varphi_n}$, and fluctuations $\epsilon_n = \varphi_n -
\varphi$ with variance $\avg{\epsilon^2} = \avg{\delta \varphi^2}$.

Substituting Eqs.~\eqref{scalar1overtn_delta}
and \eqref{scalarrnovertn_delta}
into Eq.~\eqref{Mn} and then comparing with Eq.~\eqref{Para}, one has
 \begin{align}
  e^{i\alpha} \sec \phi & = \left(\cos \varphi - i s \frac{\sin \varphi}{\cos\theta}\right)e^{ik_xl}, \label{Sa}\\
  e^{i\beta} \tan \phi & = s \sin \varphi \tan\theta  e^{-i\theta} e^{i k_x l}. \label{Sb}
\end{align}
Taking the real part of the first relation, we find the clean dispersion
for this $\delta$GSL,
\be \label{dispersion_DeltaGSL_scalar}
   \cos \mu = \cos k_xl \cos \varphi  +
   \frac{\eps}{k_xl} \sin k_xl \sin \varphi.
\ee
We recall $k_xl = [\eps^2 - l^2 k_y^2]^{1/2}=s \eps\cos\theta$.
This relation links the energy
$\eps$ to the Bloch-vector
components $K_x=\mu/l$ and $K_y = k_y$ in the bulk superlattice. The dispersion is periodic in the potential strength
$\varphi$, as discussed in detail by Barbier et al.\cite{Peeters2009_2}, and it
suffices to consider  $0\leq\varphi< 2\pi$. In the following, the
implications of disorder are assessed.

\subsubsection{Analytical Lyapunov exponent}

From Eq.~\eqref{Sb}, it becomes apparent that $\beta =  k_x
l -\theta $ is independent of $\varphi$. Consequently, $\beta'=0$ in
Eq.~\eqref{LyaF}. Furthermore, by combining both relations,
we can evaluate $\partial_\varphi(\sin\alpha/\sin\phi)$
such that the weak-disorder Lyapunov exponent finally reads
\be
    \label{LyaDelSF}
   \gamma = \frac{\avg{\delta\varphi^2}}{2} \frac{\sin^2k_xl}{\sin^2 \mu}
    \tan^2
     \theta.
\ee
This expression differs in a characteristic manner from the
corresponding result for a massive Schr\"odinger
particle:\cite{Krokhin2002}
instead of being inversely proportional to the energy, it is
proportional to $\tan^2\theta$. This has (at least) three important implications:
First, instead of diverging as $\eps^{-1}$, the weak-disorder result of Eq.~\eqref{LyaDelSF} stays valid even at low
energy $\eps$. Second,
for perpendicular incidence $\theta=0$, there is no
localization, $\gamma=0$, as required by chiral symmetry via Klein
tunneling\cite{Katsnelson2006,ZDWang2009}.
Third, the overall angular dependence as
$\tan^2 \theta$ implies that charge carriers incident with larger angles are
localized quite rapidly. Therefore, a random $\delta$GSL can act as a
\emph{directional filter}, with preferential transmission perpendicular to the
superlattice barriers.

Indeed, in the case $\varphi = 0$, i.e.,  for a purely random potential
without a regular superlattice component, Eq.~\eqref{dispersion_DeltaGSL_scalar} reduces to $\mu=k_xl$,  and the
Lyapunov exponent
\be
   \label{GV0}
 \gamma =  \frac{\avg{\delta\varphi^2}}{2} \tan^2 \theta
\ee
becomes totally independent of energy $\eps$. The simple, and sharp angular
dependence $\tan^2\theta$ realizes a disorder filter at larger
angles, allowing only particles around perpendicular incidence
$\theta=0$ to transmit ballistically.

Of course, a richer angular structure arises
via the dependence on $k_xl = s\eps \cos\theta$ and the dispersion
relation in Eq.~
\eqref{dispersion_DeltaGSL_scalar}, so that a more
detailed discussion is in order.
It is helpful to distinguish two limiting cases.
First, for $\eps\to 0$ and thus $k_xl\to0$,
Eq.~\eqref{dispersion_DeltaGSL_scalar} always has the real (hence propagating) solution
$\mu = K_xl = \varphi$. Therefore, unless $\varphi = 0,\pi$, one has $\sin k_xl/\sin\mu\to0$,  and
therefore delocalization ($\gamma\to 0$) occurs for all angles as $\eps\to 0$.
Second, for large enough $|\eps|\ge \pi$, one has $\sin k_xl=0$ for
$\eps \cos\theta_n = n\pi$ ($n\neq 0$),
while allowing $\cos\mu=\pm\cos\varphi\neq 0$. Equation~\eqref{LyaDelSF} then indicates that $\sin k_xl=0$
gives angular transmission windows $\gamma=0$ at 
\be  \label{theta_n}
\theta_n =
\arccos (n\pi/\eps).
\ee
Therefore, if the incident angle of a 2D plane wave is $\theta_n$, then the Dirac particle will be delocalized.

\begin{figure}
\centering
   \includegraphics[width=\linewidth]{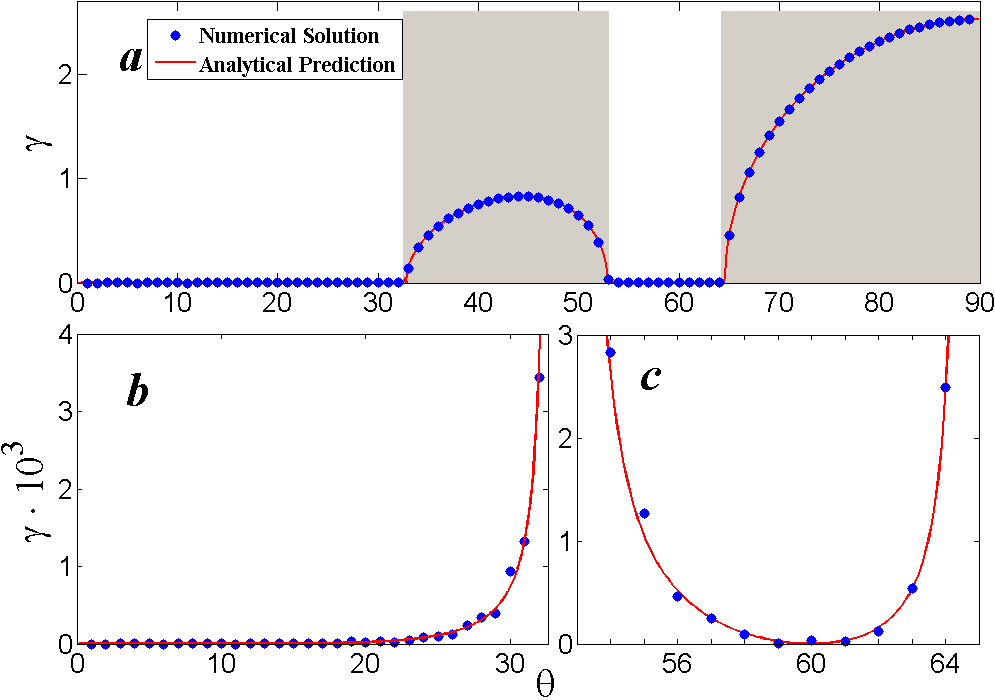}
   \caption{Lyapunov exponent $\gamma$, Eq.~\eqref{Lya}, vs.\
     angle $\theta$ at energy $\eps= El/\hbar v_F= 2\pi$ in a
     lattice of $\delta$-barriers with $\pm5\%$ fluctuations around
     the average strength $\varphi=\pi/2$.
In the grey shaded areas in panel \textbf{a}, the angle falls into a
band-gap sector, and $\gamma$ is given by Eq.~\eqref{gamma_gap}. White areas:
conduction sectors, with $\gamma$ due to disorder given by
Eq.~\eqref{LyaDelSF},
shown on a magnified scale in panels \textbf{b}, \textbf{c}.
The numerical data confirms the delocalization resonance at
$\theta_1= \arccos(\pi/\eps) = 60^{\circ}$, Eq.~\eqref{theta_n}.}
   \label{fig:DeltaS}
\end{figure}

\subsubsection{Numerical experiment}

We now turn to numerical experiments in order to check 
these predictions.
The localization Lyapunov exponents are extracted numerically by use
of Eq.~\eqref{Lya}, after first multiplying random matrices according to Eq.~\eqref{PN}.
Unless specified otherwise, we always take $N=1000$ random potential barriers and then ensemble-average
over 30 samples to reach negligible statistical error.

Figure~\ref{fig:DeltaS} compares the analytical result of
Eq.~\eqref{LyaDelSF} with the numerical data, for a varying incident
angle $\theta$
at fixed energy $\eps= 2\pi$. The
average lattice strength
is set to $\varphi = \pi/2$, and we allow
$5\%$ equiprobable fluctuations
($\avg{\delta v^2}/v^2=0.01/12$).
In the overview panel \textbf{a}, grey shading shows the intervals
where $\eps$ falls into a band gap. The
incident wave then turns into an evanescent wave, whose attenuation is
described by $\gamma$ of Eq.~\eqref{gamma_gap},
with negligible corrections due to disorder.
For the given parameters,
the band edges are located at angles $\theta_b$ solving $\cos\theta_b = \pm
\sin(\eps\cos\theta_b)$, i.e.\ $\theta_b \in \{ 32.7^\circ,
52.9^\circ, 64.6^\circ\}$. At these points, Fig.~\ref{fig:DeltaS}
shows hardly visible spikes, where the perturbative result of Eq.~\eqref{LyaDelSF} is expected
to fail.\cite{Derrida1984,Izrailev1998}
Inside the conduction intervals, shown in the magnified
view of panels \textbf{b} and \textbf{c}, the Lyapunov exponent
$\gamma$ given by Eq.~\eqref{LyaDelSF} is in excellent
agreement with numerical results.  The
delocalization resonance at $\theta_1= 60^\circ$
is confirmed, with $\gamma$ vanishing there.

\subsubsection{Exact delocalization resonance}

Interestingly, the numerical evidence suggests that the
delocalization resonance not only holds perturbatively to order
$\epsilon^2$, as predicted by Eq.~\eqref{LyaDelSF}, but instead is an
exact resonance. So we seek non-perturbative insights
by returning to the transfer matrices.
Under the resonance condition, $k_x l= \eps \cos\theta_n=n\pi$, the
free propagation phase is $e^{i k_x l}=\pm 1$. Without losing generality,
let us assume $e^{i k_x l}=1$.  The transfer matrix $M_n$ in
Eq.~\eqref{Para} then 
depends on the random variable $\varphi_n$ via 
\be
\label{M_phi}
  M_n=M(\varphi_n) = \begin{bmatrix}
             \cos \varphi_n - i s \frac{\sin \varphi_n}{\cos\theta} &  s \sin \varphi_n \tan\theta  e^{-i\theta} \\
            s \sin \varphi_n \tan\theta  e^{i\theta}&  \cos \varphi_n + i s \frac{\sin \varphi_n}{\cos\theta}
      \end{bmatrix}.
\ee
The product of transfer matrices obeys the remarkable property
\be
M(\varphi_{n})M(\varphi_{n-1}) = M(\varphi_n + \varphi_{n-1}).
\ee
Hence, the net transfer matrix across $N$ barriers is
 $P_N 
= M(\Phi_N)$
where $\Phi_N=\sum_{n=1}^N {\varphi_n}$.
As a consequence, the transmission probability 
\be
T_N = \frac{1}{\cos^2 \Phi_N + \frac{\sin^2 \Phi_N}{\cos^2 \theta}} = \frac{\cos^2\theta}{1 - \sin^2 \theta \cos^2 \Phi_N}.
\label{TN}
\ee
is bounded from below by  
$\cos^2\theta$. So for the resonance angles $\theta_n$ of
Eq.~\eqref{theta_n}, $T_N$ cannot be 
an exponentially decaying function of $N$, thus proving  $\gamma=0$.  We emphasize that this delocalization
is no longer based on a weak-disorder expansion. Rather,
it is an exact result for arbitrary disorder strength.

\subsection{Disordered square scalar potential}
\label{generalscalarpot.sec}

We return to the general case of a rectangular potential superlattice, and proceed
as previously.  With Eqs.~\eqref{scalar1overtn}
and \eqref{scalarrnovertn}
used in 
\eqref{Mn}, the comparison with Eq.~\eqref{Para} yields
 \begin{align}
  e^{i\alpha} \sec \phi & = e^{i \delta} \left(\cos \varphi - i \sin \varphi \frac{\eps v -
  \kappa^2}{\kappa ql }\right), \label{SGa}\\
  e^{i\beta} \tan \phi & = e^{i (\delta- \theta)}  \tan\theta \frac{v \sin \varphi}{s ql} . \label{SGb}
\end{align}
We denote $\kappa = lk_x = s\eps \cos\theta$ and $ql = [v^2 - 2 \eps v
+\kappa^2]^{1/2}$, as well as $\varphi = qw $. We have introduced
$\delta = k_x d $ as
the phase picked up over the distance $d=l-w$ between barriers on average.
In terms of these parameters, the dispersion relation of the clean
GSL reads\cite{Peeters2010}
\be
    \label{DisperS}
    \cos \mu=  \cos \delta \cos \varphi + \frac{\eps v-\kappa^2
    }{\kappa ql} \sin\delta \sin \varphi.
\ee

\subsubsection{Lyapunov exponent}

In the disordered case, Eq.~\eqref{SGb} fixes  $\beta = k_x d -\theta$, which
now depends on the distance $d=l-w$ between consecutive barriers,
such that $\beta'$ in Eq.~\eqref{LyaF} is finite for barriers of variable
distance $d$.
By combining Eqs.~\eqref{SGa} and \eqref{SGb}, one finds
\be
\label{FinalLya}
    \gamma=\frac{\avg{\epsilon^2}}{2} \left\{
\frac{v^2 \sin^2\varphi}{q^2l^2 \sin^2 \mu} [S']^2
 + \beta'^2\right\}
\frac{v^2 \sin^2\varphi}{q^2l^2}
\tan^2 \theta,
\ee
where $S'$ denotes the derivative of
\be \label{Svwdef}
S = \frac{ql\sin\delta\cos\varphi + k_xl \cos\delta
     \sin\varphi}{v\sin\varphi} - \frac{\eps}{k_xl}\cos\delta
\ee
with respect to the fluctuating
barrier parameter.

\begin{figure}
\centering
   \includegraphics[width=\linewidth]{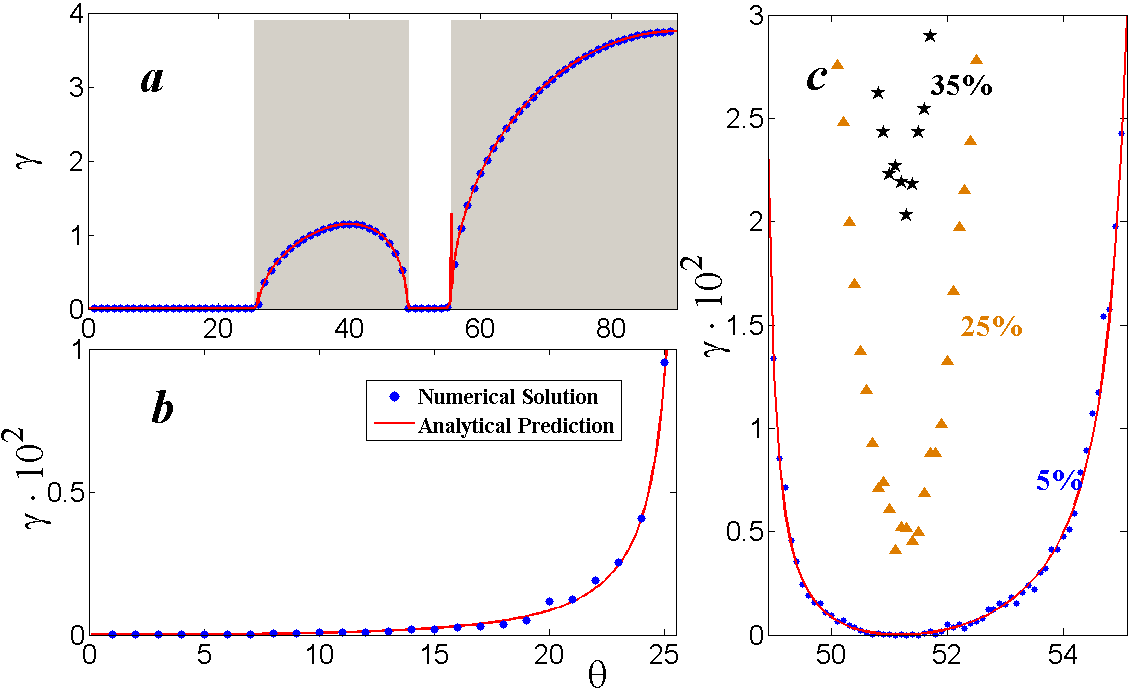}
   \caption{Lyapunov exponent $\gamma$, Eq.~\eqref{Lya}, vs.\
     angle $\theta$ at energy $\eps= El/\hbar v_F= 2\pi$ in a
     lattice of potential barriers with fixed width $w= 0.5 l$, and
     $5\%$ fluctuations around average height $v=Vl/\hbar v_F=\pi$,
     such that 
     $w v/l = \pi/2$ matches the $\delta$ barrier
     strength $\varphi$ used in Fig.~\ref{fig:DeltaS}. 
In the grey shaded areas in panel \textbf{a}, the angle points into a
band-gap direction, and $\gamma$ is given by Eq.~\eqref{gamma_gap}. White areas:
conduction sectors, with $\gamma$ due to disorder given by
Eq.~\eqref{FinalLya},
shown on a magnified scale in panels \textbf{b}, \textbf{c}.
The numerical data for stronger disorder in panel \textbf{c} shows
that the perturbative delocalization resonance at
$\theta \approx 51.5^{\circ}$ exists only for weak disorder.} 
   \label{fig:L05}
\end{figure}

The relation between the Lyapunov exponent $\gamma$ and the incidence
angle $\theta$ established by Eqs.~\eqref{FinalLya} and \eqref{Svwdef}
is rather complicated. But the overall factor $\tan^2\theta$ guarantees
absence of localization due to Klein tunneling in the forward
direction, as well as the efficient filtering of large-angle components.
To verify our analytical prediction, we plot in
Fig.~\ref{fig:L05} the Lyapunov exponent $\gamma$
as function of $\theta$ at fixed $\eps$ for barriers with randomly
varying height,
together with the data from the numerical solution. The agreement is
excellent.

\subsubsection{Approximate delocalization resonance}
\label{scalarRes.sec}
The numerical results show that there exists a delocalization resonance
$\gamma=0$ also in
this case, but at a slightly different angle, $\theta\approx
51.5^\circ$, compared to the $\delta$-barriers of
Sec.~\ref{deltascalarpot.sec}.
Let us see how this result comes about. With $\beta'=0$ (because
in Fig.~\ref{fig:L05} only the barrier height fluctuates) in
Eq.~\eqref{FinalLya}, there are two non-trivial factors that can
vanish, $\sin\varphi$ and $S'$.

First, there is the obvious candidate $\sin \varphi=0$, which is the single-barrier resonance condition $qw
=n\pi$.\cite{Katsnelson2006} But this zero is exactly cancelled by the
most singular contribution to $S'$ in Eq.~\eqref{Svwdef}, namely
$S' = q \varphi'\sin \delta /(v\sin^{2}\varphi)+O\left[(\sin\varphi)^{-1}\right]$.
Since for $\sin\varphi=0$ the dispersion relation Eq.~\eqref{DisperS} reads  $\cos\mu=\pm\cos\delta$, Eq.~\eqref{FinalLya} can be further reduced to
\be\label{gamma_res_scal}
     \gamma=\frac{\avg{\delta v^2}}{2}
    \frac{w^2 v^2 (v-\eps)^2}{l^6q^4}
\tan^2 \theta.
\ee
This expression could be thought to vanish for $v=\eps$, i.e.,  when the energy equals the mean potential height.
However, $v=\eps$ implies $q=i k_y $, which is impossible because it
contradicts the initially assumed resonance condition $qw  =n\pi$.

Therefore, $S'=0$ must be responsible for the observed delocalization resonance
$\gamma=0$. In general, the equation $S'=0$ is too complicated to admit
an analytical solution,
but the resonance angles $\theta_n$  can be found numerically. For the
 present parameters it is the
resonance angle $\theta_1$ that is observed in Fig.~\ref{fig:L05}. In contrast to
the case of $\delta$-barriers, though, this resonance is not
exact. In Fig.~\ref{fig:L05}c, numerical results for stronger disorder
show a deviation from $\gamma=0$, thus indicating the absence
of a true delocalization resonance.

We note that for a purely random
potential, $v=0$ and $\epsilon_n = v_n$,
the Lyapunov exponent Eq.~\eqref{FinalLya} reduces to
\be
\label{FinalLyaV0}
    \gamma 
    = \frac{\avg{\delta v^2}}{2} \frac{\sin^2(k_x w)}{k_x^2l^2} \tan^2 \theta,
\ee
Now the single-barrier resonance condition
$k_xw = n\pi$ does lead to $\gamma=0$. Consistently, the limit $v\to0$ of
Eq.~\eqref{gamma_res_scal} vanishes. Here, to lowest order in
$\epsilon=\delta v$, the wavevector inside the barrier is
$k_x$, and the resonance condition can be satisfied everywhere.
But it needs to be emphasized that also this result holds only for weak disorder, and hence the Lyapunov
exponent is not absolutely zero due to higher-order terms of $\epsilon$.

\section{Vector potential}
\label{vectorpot.sec}

This section  parallels  the
previous one, with results pertaining to disordered vector-potential GSLs, as introduced in Eq.~\eqref{vectorpot}.
The single-barrier reflection and transmission amplitudes $r$ and $t$
are
\cite{Neto2009}
\begin{align}
   \frac{1}{t} & = e^{i w k_x} \left(\cos \varphi - i \sin \varphi \frac{u \sin\theta + \eps \cos^2\theta}{\tilde q l \cos\theta}\right), \label{VPt} \\
   \frac{r}{t} & = e^{i w k_x} e^{i\theta} \sec\theta \frac{u \sin \varphi}{\tilde q l} . \label{VPrt}
\end{align}
Here $\eps = E l/\hbar v_F = s|\bk|l $ and  $u = e A l/\hbar c$ are energy and
barrier height expressed in lattice units. Besides,
$\varphi = \tilde q w $ is the phase picked
up by the plane wave with wavevector $\tilde q =l^{-1}
[\eps^2 - (l k_y - u)^2]^{1/2}$  
across the potential barrier.
The variable $\tilde q$ differs from the wavevector $q$ in the previous scalar potential case.
In particular, $\tilde q$ can be imaginary if $u$ is large,
leading to bound states inside a barrier.\cite{Neto2009}


\subsection{Amplitude-disordered  delta vector potential}
\label{deltavectorpot.sec}

Very narrow and high potentials barriers 
\ie, $k_x w \ll 1 $ and $u \gg \eps $, realize a vector $\delta$GSL.
In the limit $w\to0$ and $u\to\infty$ at fixed $uw/l = \varphi$,
one has $\tilde q\to i u/l$, and the reflection and transmission coefficients
in Eqs.~\eqref{VPt} and \eqref{VPrt} reduce to
\begin{align}
  \frac{1}{t}& = \cosh \varphi - i \sinh \varphi \tan \theta, \label{VPt_delta}\\
  \frac{r}{t}& = e^{i\theta} \sec\theta \sinh \varphi. \label{VPrt_delta}
\end{align}
The fluctuating phase $\varphi=uw/l$ describes randomness in both width $w$ and
height $u$.
We assume a distribution with mean
$\varphi=\avg{\varphi_n}$, and small fluctuations $\epsilon_n = \varphi_n -
\varphi$ with variance $\avg{\epsilon^2} = \avg{\delta \varphi^2}$.

Substituting Eqs.~\eqref{VPt_delta} and \eqref{VPrt_delta} into Eq.~\eqref{Mn} and comparing with Eq.~\eqref{Para},
one has
\begin{align}
  e^{i\alpha} \sec \phi & = (\cosh \varphi + i \sinh \varphi \tan \theta)e^{ik_xl}, \label{Va}\\
  e^{i\beta} \tan \phi & = - \sinh \varphi \sec\theta  e^{-i\theta} e^{i k_x l}. \label{Vb}
\end{align}
The clean dispersion \eqref{Dispersion0} for the vector $\delta$GSL is found
by taking the real part of the first relation:
\be
   \label{DisV}
    \cos \mu = \cosh \varphi \cos \kappa - \tan \theta \sinh \varphi \sin \kappa.
\ee
where $\kappa = k_xl = [\eps^2 - l^2 k_y^2]^{1/2}$.

\subsubsection{Lyapunov exponent}

For regularly spaced potentials, $\beta'=0$ in
Eq.~\eqref{LyaF}. Using Eqs.~\eqref{Va} and \eqref{Vb}
to evaluate $\partial_\varphi(\sin\alpha/\sin\phi)$,
we find the weak-disorder Lyapunov exponent
\be
   \label{LyaDeltaV}
   \gamma = \frac{\avg{\delta \varphi^2}}{2} \frac{\sin^2 k_xl}{\sin^2 \mu} \sec^2 \theta.
\ee
This expression ressembles much the scalar $\delta$GSL result \eqref{LyaDelSF},
except that the $\tan^2\theta$ factor is replaced by $\sec^2 \theta$.
Therefore, vector $\delta$GSL and scalar $\delta$GSL share (at least) one
interesting feature: Instead of diverging as $\eps^{-1}$ in the Schr\"odinger case,
the weak-disorder prediction of the localization length stays valid even at low
energy $\eps$. On the other hand, because $\sec^2 \theta=1$ at $\theta=0$,
there is no reason to expect delocalized solutions close to
perpendicular incidence on general grounds. 

\subsubsection{Absence of delocalization resonances}

For $\varphi =0$, representing a random vector potential with zero
mean, Eq.~\eqref{LyaDeltaV} reduces to the energy-independent expression
$\gamma =  \frac{1}{2}\avg{\delta\varphi^2}\sec^2 \theta$.
The angular dependence 
$\sec^2\theta$ differs from Eq.~\eqref{GV0} for scalar $\delta$GSL in
that localization stays finite even at perpendicular incidence
$\theta=0$, but becomes just as strong at grazing incidence
$\theta\to\pi/2$ where
$\sec\theta\approx\tan\theta$.

For the general situation with $\varphi\neq 0$, we analyze two representative cases.
Consider first the low-energy limit $\eps\to 0$ and thus $\kappa\to0$. Then the dispersion
Eq.~\eqref{DisV} reads
$\cos\mu =\cosh \varphi$, which requires an imaginary $\mu$
and hence describes a non-propagating solution inside the band gap.
As such, the vector $\delta$GSL acts as an insulator for small $\eps$
and arbitrary incidence angle $\theta$.

Next we turn to cases with sufficiently large $|\eps|\ge \pi$. If $\eps \cos\theta_n = n\pi$ ($n\neq 0$),
one has $\sin\kappa=0$ and hence $\gamma=0$. Note that under this condition, $|{\cos\mu}|=\cosh\varphi$,
which again implies a non-propagating solution. So here $\gamma=0$ merely
indicates that the disorder-induced correction to the decay exponent of the evanescent wave is zero. Putting all the above
considerations together, it appears that the localization behavior in vector $\delta$GSLs is not as rich as
in scalar $\delta$GSLs.

\begin{figure}
\centering
   \includegraphics[width=\linewidth]{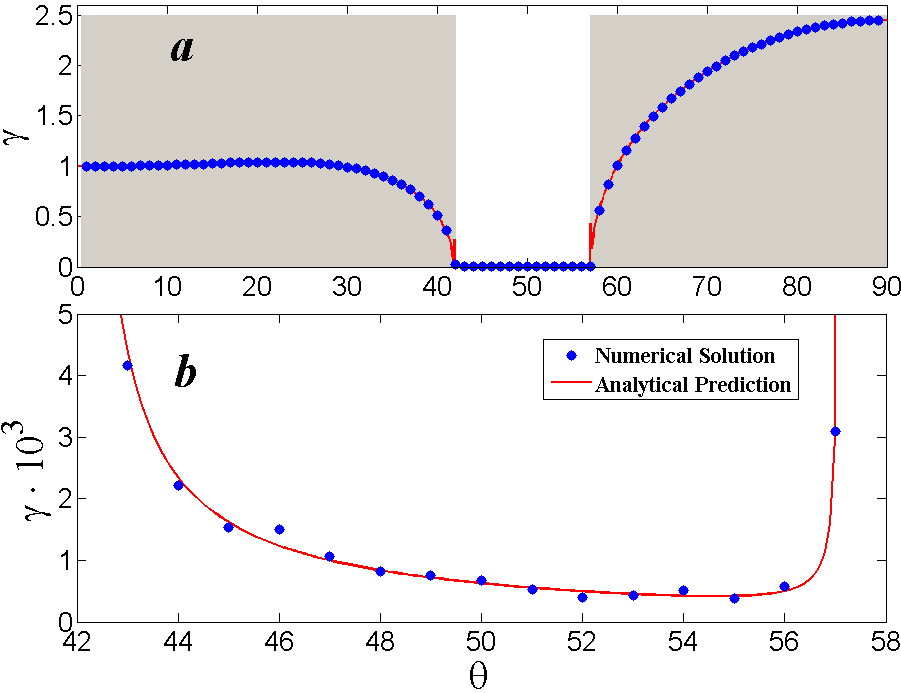}
   \caption{Lyapunov exponent $\gamma$, Eq.~\eqref{Lya}, vs. incident angle $\theta$, for
   a vector $\delta$GSL, with average lattice strength $\varphi=1$,
   energy $\eps=2\pi$, and
   disorder modeled by $\pm 5\%$ fluctuations around $\varphi$.  In the grey shaded areas
   in panel \textbf{a}, the incident angle falls into a band-gap direction. The white area is the
    conduction sector, in which the anaytical Lyapunov exponent is given by Eq.~\eqref{LyaDeltaV}, as
    shown in Panel \textbf{b} on a larger scale.}
   \label{fig:DeltaV}
\end{figure}

Figure~\ref{fig:DeltaV} compares the analytical prediction
Eq.~\eqref{LyaDeltaV} with numerical results, for varying incidence
angle $\theta$ at fixed energy $\eps= 2\pi$. The
average lattice strength is  $\varphi = 1$, and disorder is
modeled by $5\%$ equiprobable fluctuations around $\varphi$.
In the overview panel \textbf{a}, band-gap regimes are grey shaded.
The band edges lie at the angles $\theta_b \in \{ 42^\circ,
57^\circ\}$. Exactly at these points, abnormal spikes are seen
in panel \textbf{a}, signaling the expected failure of Eq.~\eqref{LyaDeltaV}.\cite{Derrida1984,Izrailev1998}
Inside the conduction band, shown on a magnified scale in panel \textbf{b}, the agreement between
theory and numerics is excellent. In particular, no delocalization resonance is seen, as analyzed above.

\subsection{Disordered square vector potential}
\label{generalvectorpot.sec}

The general case of a disordered rectangular vector potential is somewhat
more complicated and considerably richer in physics.
Following the same procedure as for scalar GSLs, the first step is to connect
the transfer-matrix parameters $\alpha$, $\beta$ and $\phi$ to the GSL parameters $w$, $u$, and $l$, as well as the Dirac-particle quantum numbers $\eps,\theta,s$.  For that purpose we
use Eqs.~\eqref{VPt}, \eqref{VPrt}, \eqref{Mn}, and \eqref{Para} to
obtain
 \begin{align}
  e^{i\alpha} \sec \phi & = e^{i\delta} \left(\cos \varphi + i \sin
    \varphi \frac{\kappa^2 + u \eps \sin\theta}{l\tilde q \kappa}\right), \label{VPa}\\
  e^{i\beta} \tan \phi & = -e^{i(\delta-\theta)} \frac{u
    \sin \varphi}{l\tilde q}  \sec\theta. \label{VPb}
\end{align}
Here $\kappa = lk_x = \eps \cos\theta$, $l\tilde q = [\eps^2 - (l k_y - u)^2]^{1/2}$,
and $\varphi = \tilde q w$.
$\delta = k_x (l-w)$ is 
the phase picked up
between neighboring barriers.
In terms of these parameters, the dispersion relation of a clean
GSL becomes
\be
    \label{DisperV}
    \cos \mu= \cos \delta \cos \varphi - \frac{\kappa^2 + u
      \eps\sin\theta}{l\tilde q\kappa} \sin \delta \sin \varphi.
\ee

\subsubsection{Lyapunov exponent}

Using Eqs.~\eqref{VPa} and \eqref{VPb}, we can apply our general result Eq.~\eqref{LyaF} once again, leading to
\be
\label{FinalLyaV}
    \gamma=\frac{\avg{\epsilon^2}}{2} \left\{
\frac{u^2 \sin^2\varphi}{l^2\tilde q^2 \sin^2 \mu} [\tilde S']^2
 + \beta'^2\right\}
\frac{u^2 \sin^2\varphi}{l^2\tilde q^2}
\sec^2 \theta ,
\ee
where $\tilde S'$ denotes the derivative of
\be \label{SvwdefV}
\tilde S = \frac{l\tilde q\sin\delta\cos\varphi + \kappa \cos\delta
     \sin\varphi}{u\sin\varphi} + \frac{\eps}{\kappa}\cos\delta\sin\theta
\ee
with respect to the fluctuating
barrier parameter.
In contrast to the scalar potential with overall $\tan^2\theta$
dependence, the factor $\sec^2\theta$  in Eq.~\eqref{FinalLyaV}
does not lead to a simple
delocalization resonance at perpendicular incidence, just as for the vector $\delta$GSL of Sec.~\ref{deltavectorpot.sec}.

\begin{figure}
\centering
   \includegraphics[width=\linewidth]{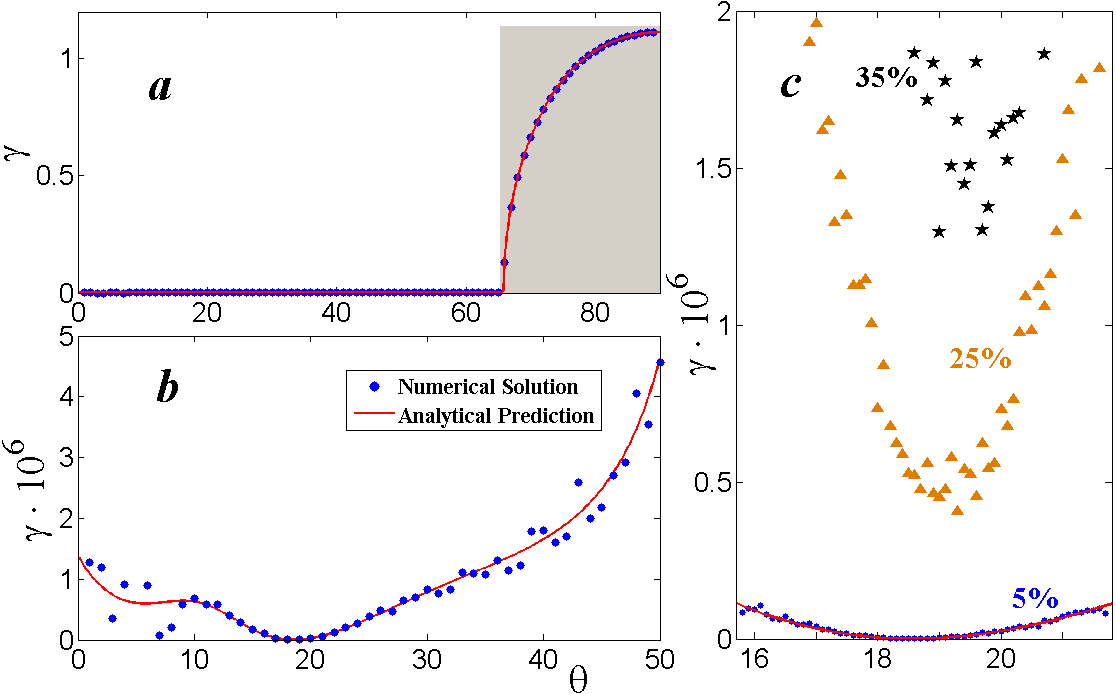}
   \caption{Lyapunov exponent $\gamma$, Eq.~\eqref{Lya}, vs. incident angle $\theta$,
   at energy $\eps=2\pi$ for a disordered lattice of rectangular vector potentials.
   Disorder is modeled by $\pm 5\%$ fluctuations around an average barrier
   height $u=2$, while periodicity $l$ and barrier width
   $w=0.5l$ are fixed, such that 
$w u/l = 1$ matches the $\delta$ barrier strength $\varphi$ used
   in Fig.~\ref{fig:DeltaV}. 
  The grey shaded area
   in panel \textbf{a} indicates a band gap and the white area
   indicates conducting solutions.
    Panels \textbf{b} and \textbf{c} show details of the conduction sector.
   An approximate delocalization resonance appears around $\theta=18.5^{\circ}$, for both analytical and numerical results.
   The numerical data for stronger disorder in Panel \textbf{c} proves the departure from the weak-disorder
   resonance. }
   \label{fig:L05V2}
\end{figure}

Our numerical data confirm these predictions, as seen in
Fig.~\ref{fig:L05V2}. The statistical fluctuations in
Fig.~\ref{fig:L05V2}b appear larger than before because
for the present parameters, the Lyapunov exponent $\gamma$ is
extremely small. 

\subsubsection{Approximate delocalization resonance}

Figure \ref{fig:L05V2} also reveals 
a delocalization resonance $\gamma=0$ at $\theta\approx
18.5^\circ$, all the more remarkable because no such resonance occurs in the $\delta$-barrier limit
of Sec.~\ref{deltavectorpot.sec}.
In order to explain this analytically,
we return to Eq.~\eqref{FinalLyaV}. First of all, for the amplitude
randomness studied in Fig.~\ref{fig:L05V2},
$\beta'=0$. Then, $\gamma=0$ at $u\neq 0$ implies $\sin\varphi=0$ or
$S'=0$.

Let us begin by analyzing the case $\sin \varphi=0$, which is
equivalent to the barrier resonance condition
$\tilde q w= (w/l)[\eps^2 - (\eps\sin\theta -u)^2]^{1/2} = n\pi$.
To leading  order in  $1/\sin\varphi$,
we find $|{S'}| = |q \varphi' \sin\delta / (u \sin^2\varphi)|$ from
Eq.~\eqref{SvwdefV}. So 
the $\sin^4\varphi$ factors cancel in
Eq.~\eqref{FinalLyaV}, which reduces to
\be
\label{vec_res}
\gamma=\frac{\avg{\delta u^2}}{2} \frac{w^2 u^2 (u-lk_y)^2}{l^6\tilde q^4} \sec^2 \theta.
\ee
This expression vanishes (remember $u\neq 0$)
for  $u=lk_y$, which is equivalent to $u=\eps\sin\theta$.
Together with the barrier resonance condition, this fixes $\eps_n = n\pi l/w$.
Therefore, resonances should
occur whenever
\be\label{tildethetan}
\tilde \theta_n = \arcsin (u/\eps_n)
\ee
For the parameters of
Fig.~\ref{fig:L05V2} ($u=2$, $\eps=2\pi$, and $w=l/2$),
Eq.~\eqref{tildethetan} predicts a resonance at $\tilde \theta_1\approx
18.6^{\circ}$, in perfect agreement with the data in
Fig.~\ref{fig:L05V2}\textbf{c}. As shown by the data for stronger
disorder, the delocalization resonance only holds to lowest order of the
weak-disorder expansion.  

Are there other delocalization resonances caused by $S'=0$?
A direct answer is difficult on account of the rather complex
expression for $S'$.
Numerically,  we have scanned the values of $S'$ and find
that when $S'$ is zero, the associated solution falls inside a band gap. This being the case,
the $S'=0$ condition does not produce new delocalization resonances,
in marked difference to the scalar GSLs studied in
Sec.~\ref{scalarRes.sec}.

\section{Wave Packet Dynamics: disorder-induced filtering}
\label{wavepackeet.sec}

Our analytical results have revealed an interesting functional dependence of the localization length upon the incident angle
of charge carriers. In particular, the Lyapunov exponent
$\gamma=l/l_\text{loc}$ of a scalar GSL is proportional to
$\tan^2\theta$.  This factor indicates a strong angular dependence of
disordered-induced
localization: the localization length diverges for small $\theta$ and quickly decreases as $\theta$ increases.  Certainly,
for $\theta$ too close to $\theta=\pi/2$, an infinite Lyapunov
exponent or vanishing localization length
is an artifact of weak-disorder perturbation theory.
With this clarified, it is nevertheless clear that
scattering waves with
larger $\theta$ tend to be much more localized than
those with small $\theta$. And wave components with localization length shorter than
the GSL sample will not contribute to the conductance. This realizes a
filtering effect due to disorder.
The main goal of the present, comparatively short section is to
confirm this effect by a direct dynamical simulation of wave-packet
transmission across a scalar GSL, both with and without
disorder. 

\begin{figure}
\centering
   \includegraphics[width=\linewidth]{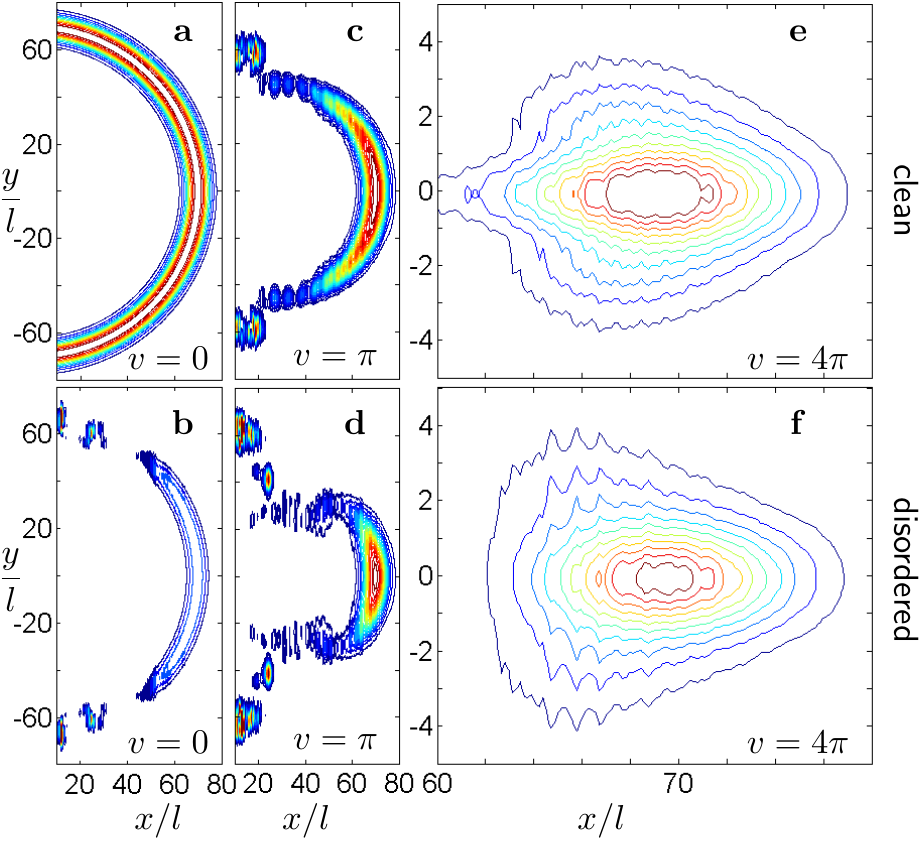}
   \caption{
Contour plot of the probability density, from 2D wave-packet dynamics
simulations, for various
amplitudes $v=Vl/\hbar v_F$ of clean (upper row) and
amplitude-disordered (lower row) scalar GSL potentials $V(x)$,
Eq.~\eqref{scalarpot}. 
The time evolution samples all incidence angles $\theta$ at once,
starting with an isotropic wave packet centered around energy
$\eps=El/\hbar v_F=2\pi$ (see text).  
   Comparison between panels \textbf{a}-\textbf{b}, and \textbf{c}-\textbf{d} demonstrates disorder-induced
   filtering: since wave-packet components at larger angle
   $\theta$ have a shorter localization length, they cannot contribute
   to propagation in $x$ direction, and the transmitted part of the wave packet appears more focused.
   The GSL potential in panels \textbf{e}-\textbf{f} is sufficiently strong to induce the wave-packet
   collimation that accompanies the emergence of new Dirac
   cones. Panel \textbf{f} shows that disorder has rather little effect on collimation.}
   \label{fig:AllTogether}
\end{figure}

Figure \ref{fig:AllTogether} shows the result of a numerical solution of the time-dependent Dirac
equation with Hamiltonian \eqref{First} and a scalar GSL potential,
Eq.~\eqref{scalarpot}, with symmetric barrier width $w=0.5l$ filling the
half-space $x>10l$.   
In order to sample all incidence angles at once, we
choose as initial condition an isotropic
wave packet with momentum components $\Psi(p) \propto
\exp\{-(|\bp|-p_0)^2/(2\Delta p^2)\}$ centered on the radial value $p_0 = 2\pi
\hbar/l$ with spread $\Delta p=0.2 \hbar/l$; the wave packet's central energy therefore is $\eps=2\pi$ in
lattice units. 
In Fig.~\ref{fig:AllTogether}, we plot the
probability density at time $t=70l/v_F$; in some cases, a substantial part of the wave packet is reflected into
the half-space $x<10l$ (not shown). The upper row shows the results for clean GSLs of different strengths,
whereas the lower row shows the results for a single realization of
the corresponding disordered GSLs with fluctuating potential heights.

Panels \textbf{a} and \textbf{b} compare a pristine graphene sheet to 
a purely amplitude-disordered scalar GSL with zero mean potential
strength and equiprobable fluctuations $\delta v\in[-1,1]$.  
Whereas the clean substrate allows for isotropic
propagation, in the disordered GSL the larger-angle components are
localized more strongly, as expressed by the $\tan^2\theta$-behavior of
the Lyapunov exponent,
Eq.~\eqref{FinalLyaV0}. 
Consequently, the propagating part of the wave packet is
concentrated around the forward direction $\theta=0$, thus supporting
our filtering conjecture above. 

Panels \textbf{c} and \textbf{d} compare again the clean and
disordered situation, now in presence of a GSL with finite strength
$v=\pi$, with the same lattice geometry and energy as used for
Fig.~\ref{fig:L05}, but relatively strong amplitude fluctuations of 
$\pm 30\%$. A strong filtering effect analogous to panel \textbf{b} is
observed, 
where the largest part of the transmitted probability density
is concentrated in the forward direction $\theta=0$, 
as expressed by the overall $\tan^2\theta$-behavior of
the Lyapunov exponent, Eq.~\eqref{FinalLya}. 

The wave propagation in the clean GSL of panel \textbf{c} is quite isotropic, because
the associated dispersion relation is almost isotropic for the parameters chosen.
If, however, the potential strength of a scalar GSL is greater than a certain critical value,
new Dirac points emerge.\cite{Louie2011}
The resulting, strongly anisotropic dispersion
relation then collimates the wave packet.\cite{Louie2008_2}  This is shown in panel \textbf{e},
where the potential strength $v=4\pi$ makes the wave packet
stay sharply focused in the forward direction.  We have investigated whether this collimation effect is robust against
disorder.  Panel \textbf{f} shows the effect of $10\%$ 
fluctuations in potential strength. The collimation is
seen to survive, with hardly noticeable disorder effects.
A quantitative analysis is difficult because the new Dirac points
appear at band edges where the weak-disorder expansion we have used
fails. Instead, one could possibly adapt the appropriate singular-point
expansions \cite{Derrida1984,Izrailev1998} to the Dirac-GSL problem,
which is a research program beyond the scope of the present work.
Here, we conclude that
disorder-induced filtering can coexist with band-structure
collimation.

\section{Concluding Remarks }

Drawing on a general weak-disorder expansion,
we have derived the  Lyapunov exponent (inverse localization length) of various
1D disordered GSLs modeled by random delta or rectangular potentials,
both for scalar and vector potentials.  The analytical
results have been thoroughly checked by numerical experiments.  We
emphasize that, though the GSL is assumed to be 1D, the physics is far
more complicated than for a conventional 1D scattering problem due to the intrinsic coupling between the
translational motion and the spinor degree of freedom.  One important complication we have predicted is the strong
dependence of the localization length on the incident angle of the charge carriers injected to a GSL. To our knowledge,
this is the first time that a complete theoretical picture of 
this incident-angle dependence is obtained.
We have also
proposed to exploit such angular dependence of the localization length to turn disorder into good use, namely, a possible
disorder-assisted filtering effect.  
Considering that large-size GSLs may be manufactured in the near future, our theoretical
results offer a quantitative tool to analyze and predict disorder effects in GSLs.

Our analytical and numerical results also provide evidence for
intriguing delocalization resonances: 
Along specific incident angles,  the localization exponent 
can be identically zero, or at least approach zero for weak disorder.  Both scalar and vector GSLs admit
delocalization resonances in the conduction band, but for opposite
 reasons: scalar potentials can have an approximate, weak-disorder resonance because
 a complex term has zero solutions [i.e., $S'=0$, see Eq.~(\ref{FinalLya}) and (\ref{Svwdef})],
 whereas vector potentials have an approximate resonance because of a simple barrier
 resonance condition [$\sin\varphi=0$, see Eq.~(\ref{FinalLyaV})].  Moreover, the corresponding $\delta$-limits of scalar and vector GSLs
 show very distinct features: the scalar $\delta$GSL admits an exact delocalization by virtue
 of an inter-peak resonance, whereas the vector $\delta$GSL has no
 resonance at all in the conduction band. In all cases, it is
 important to realize that 
whenever numerical or laboratory experiments are performed with finite-size samples, a lowest-order vanishing
Lyapunov exponent can very well appear as a rather sharp mobility
jump, which signals an effective delocalization across the sample.\cite{Izrailev2001,Krokhin2002,Lugan2009}

In the context of 2D GSLs, a recent study \cite{Aihua2010} cautioned that lattice constants less than
10 nm may induce inter-valley scattering or sublattice symmetry breaking, either of which may
lead to a band gap and hence break the linear dispersion relation of the charge carriers.
The implication of this important finding for our work is twofold.  First, to directly apply
our theoretical results based on a linear dispersion relation, it is safer to
consider GSLs with lattice constants larger than 10 nm or with a
potential preserving the symmetry between different Dirac points or between different sublattices.  Second,
as a possible extension of this work, one may now also apply our main theoretical tool here to
investigate how a disordered GSL with a sufficiently small lattice constant may generate a novel physical situation, where
charge carriers possess disordered mass as a consequence of inter-valley scattering or sublattice symmetry breaking.

\acknowledgments  J.G.\ is grateful to Prof.\ Chun Zhang for stimulating discussions on graphene superlattices and
for providing several useful references on this topic. C.M.\
acknowledges helpful correspondence with Felix Izrailev.  

\appendix

\section{Details of weak-disorder expansion}
\label{lyap.apx}
This appendix provides some details of the
analytical calculation leading to the weak-disorder Lyapunov exponent
given by Eq.~\eqref{LyaM} and Eq.~\eqref{LyaF}.

\subsection{Absence of mixed-fluctuation terms}
\label{lyap1.apx}

The starting point is Eq.~\eqref{tildePN}, where
$\lambda_\pm=e^{\pm i\mu}$, $\mu\in\R$,
describes a propagating solution in the clean GSL. A Taylor expansion to
quadratic order in the fluctuations $\epsilon_n$ leads to
\begin{align}
\frac{1}{2N}\ln|(P_N)_{11}|^2  = & \Re\left\{\frac{\tilde
    M'_{11}}{\lambda_+}\right\}  \frac{1}{N}\sum_{n=1}^N \epsilon_n \label{PNappdx}\\
 & +  \Re \left \{ \frac{\tilde M''_{11}}{\lambda_+} -
    \frac{\tilde M'^2_{11}}{\lambda^2_+} \right \} \frac{1}{N}\sum_{n=1}^N
    \frac{\epsilon^2_n}{2} \nonumber \\
& + \Re \left\{\tilde M'_{12} \tilde M'_{21} \frac{1}{N} \sum_{n<m } \lambda_+^{2(n-m)} \epsilon_n \epsilon_m \right \}. \nonumber
\end{align}
First, we justify that the last line only gives a negligible contribution under the ensemble average.
In terms of the complex random variable $z_n= \epsilon_n\lambda_+^{2n}
= \epsilon_n e^{2in\mu}$,
the double sum rewrites
\be\label{dblsum}
 \frac{1}{N}\sum_{n<m } z_n z_m^* =\frac{1}{2N}\Bigg|\sum_{n=1}^N z_n\Bigg|^2 -
 \frac{1}{2N}\sum_{n=1}^N |z_n|^2
\ee
In the second term, we recognize the variance
$\overline{|z|^2} = \avg{\epsilon^2}=:\sigma^2$ in the limit $N\to\infty$. The
whole expression   \eqref{dblsum}  can be written as  $\sigma^2(|y_i|^2
- 1 )/2$, where
the random variable $ y_i \equiv \sum_{n=1}^N
z^{(i)}_n/(\sqrt{N}\sigma)$
fluctuates as samples $i$ are drawn from the ensemble.
Now, according to the Berry-Esseen theorem, in the limit $N\to\infty$
the probability distribution of $|y|$ converges to the standard normal
distribution,
with unit variance $\avg{|y|^2} =1$. As a consequence,
$\sigma^2(\avg{|y|^2} - 1)=0$, such that
the whole expression \eqref{dblsum} gives zero
contribution after the ensemble average.

Then, the vanishing fluctuation mean \eqref{avgeps} makes also the first line in \eqref{PNappdx}
vanish. Thus, only the variance \eqref{vareps}
in the second line contributes to
\be\label{LyaM.apx}
  \gamma = \lim_{N\to\infty} \frac{ \ln |(P_N)_{11}|^2}{2N} = \frac{\avg{\epsilon^2}}{2}
\Re\left\{ \frac{\tilde M''_{11}}{\lambda_+} - \frac{\big(\tilde M'_{11}\big)^2}{\lambda_+^2}
\right\},
\ee
which is the result stated as Eq.~\eqref{LyaM}.


\subsection{Diagonalization procedure }

\label{lyap2.apx}

As the last task, we need to express the matrix elements $\tilde M'_{11}$
and $\tilde M''_{11}$ of the diagonal representation through  the transfer-matrix parameters
$\{\alpha,\beta,\phi\}$ as defined in \eqref{Para}.
In an intermediate step, we parameterize the transfer matrix as
\be
    \label{MatA}
    M = \begin{pmatrix}a & b \\
    b^* & a^*\end{pmatrix},
\ee
where
\begin{align}
a & = e^{i\alpha}\sec\phi, \\
b &= e^{i\beta}\tan\phi,
\end{align}
have to satisfy the constraint $\det M = |a|^2-|b|^2=1$. The diagonal
representation $ \tilde M=\mathrm{diag}(\lambda_+,\lambda_-) =
P^{-1} M P$ is attained by a basis transformation with
\begin{align}
  P  & =  \begin{pmatrix} b & b \\
          \lambda_+ - a & \lambda_- -a
          \end{pmatrix}, \label{Paa}\\
 P^{-1} &=\frac{1}{b(\lambda_--\lambda_+)}
         \begin{pmatrix} \lambda_- -a & -b \\
                     -(\lambda_+ - a) & b
         \end{pmatrix} \label{Pbb}.
\end{align}
Again, we assume that the eigenvalues $\lambda_\pm = e^{\pm i \mu}$
form  a complex conjugate ($\mu\in \R$),
and non-degenerate ($\mu\neq n\pi$ for all $n\in\Z$) pair
since we seek the Lyapunov exponent of inside-conduction-band solutions.

Elementary algebra leads to
\be
  \tilde M'_{11}  = (P^{-1} M'P)_{11} = \frac{2 \lambda_+ \Re \{a'\} }{\lambda_+ -
    \lambda_-},
\label{B11}
\ee
where the useful identities $\lambda_+ + \lambda_- = a+a^*$ and
$(\det M)' = 2 \Re\{a^*a'-b^*b'\}=0$ have been employed.
Furthermore, $\Re\{a\}= \cos\mu$ entails $\Re\{a'\} = - \mu'\sin\mu$,
and since $\lambda_+-\lambda_-= 2i\sin\mu$, Eq.~\eqref{B11} implies
\be
   \label{1st_term}
-   \left(\tilde M'_{11}/\lambda_+ \right)^2 =  \frac{(\Re \{a'\})^2}{\sin^2 \mu} = \mu'^2,
\ee
which is the second term needed in Eq.~\eqref{LyaM.apx}.

Proceeding similarly, one finds for the first term
\be
\frac{\tilde M''_{11}}{\lambda_+} = \frac{\Re\{a''\}}{i\sin \mu} +
\frac{\Re\{a^* a'' - b^* b''\}}{-i\sin\mu} (\cos\mu - i\sin \mu).
\ee
We only need its real part,
\begin{align}
    \label{2nd_term}
    \Re \left\{ \lambda_+^{-1}\tilde M''_{11} \right\}
    & = \Re\{a^* a'' - b^* b''\} =  |b'|^2 -|a'|^2 \nonumber\\
    & = (\phi'^2-\alpha'^2) \sec^2 \phi  + \beta'^2 \tan^2 \phi .
\end{align}
Substituting Eq.~\eqref{1st_term} and \eqref{2nd_term} into Eq.~\eqref{LyaM.apx}, we have
\be
\label{Lya3}
\gamma = \frac{\avg{\epsilon^2}}{2} \left\{ \mu'^2 +(\phi'^2 - \alpha'^2) \sec^2 \phi  +\beta'^2\tan^2 \phi \right\},
\ee
Further algebraic manipulations lead to the identity
\be
\mu'^2 + (\phi'^2 - \alpha'^2) \sec^2 \phi  = \frac{\tan^4
\phi}{1-\sec^2 \phi \cos^2 \alpha}\left[\left(\frac{\sin\alpha}{\sin\phi}\right)'\right]^2,
\ee
which then results in the final expression Eq.~\eqref{LyaF} for the
weak-disorder Lyapunov exponent.



\begin{thebibliography}{99}
     \bibitem{Weiss1958}J. C. Slonczewski and P. R. Weiss, Phys. Rev. {\bf 109}, 272 (1958).
     \bibitem{Semenoff1984}G. W. Semenoff, Phys. Rev. Lett. {\bf 53},
       2449 (1984).  
     \bibitem{Haldane1988}F. D. M. Haldane, Phys. Rev. Lett. {\bf 61}, 2015 (1988).
     \bibitem{Geim2009} A.~H.~Castro Neto, F.~Guinea, N.~M.~R.~Peres,
       K.~S.~Novoselov, and A.~K.~Geim, Rev.\ Mod.\ Phys.\ {\bf 81}, 109 (2009).
     \bibitem{Sharapov2005}V. P. Gusynin and S. G. Sharapov, Phys. Rev. B {\bf 71}, 125124 (2005).
     \bibitem{Kim2006}M. S. Purewal, Y. Zhang and P. Kim, Phys. Status Solidi B {\bf 243}, 3418 (2006).
     \bibitem{Klein1929}O. Klein, Z. Phys. {\bf 53}, 157 (1929).
     \bibitem{Katsnelson2006} M. I. Katsnelson, K. S. Novoselov, and A. K. Geim, Nat. Phys. {\bf 2}, 620 (2006).
     \bibitem{Zitter}J. Schliemann, D. Loss, and R. M. Westervelt, Phys. Rev. Lett.
     {\bf 94}, 206801 (2005); W. Zawadzki, Phys. Rev. B {\bf 72}, 085217 (2005);
     R. Winkler, U. Zulicke, and J. Bolte, \textit{ibid}. {\bf 75}, 205314 (2007).
     \bibitem{Ohberg2008} G.~Juzeliunas, J.~Ruseckas, M.~Lindberg, L.~Santos, and P.~\"{o}hberg, Phys. Rev. A {\bf 77}, 011802 (2008).
     \bibitem{Lee2009}K.~L.~Lee, B. Gr\'{e}maud, R.~Han, B.-G.~Englert, and C.~Miniatura, Phys. Rev. A {\bf 80}, 043411 (2009).
      \bibitem{Oh2010}Q. Zhang, J. B. Gong, and C. H. Oh, Phys. Rev. A {\bf 81}, 023608 (2010).
     \bibitem{Pachos2011}E. Alba, X. Fernandez-Gonzalvo, J. Mur-Petit, J. J. Garcia-Ripoll, and J. K. Pachos, arXiv:1107.3673v1 (2011).
      \bibitem{Alexandrov2011}S. E. Savel'ev and A. S. Alexandrov, arXiv:1103.5983v1 (2011).
      \bibitem{Lamata2011} L Lamata, J. Casanova, R. Gerritsma, C. F. Roos, J. J. Garc\'{i}a-Ripoll, and E. Solano, New J. Phys. {\bf 13}, 095003 (2011).

      \bibitem{Zawadzki2011} W. Zawadzki and T. M. Rusin, J. Phys.: Condens. Matter {\bf 23} (2011) 143201.
       \bibitem{Unanyan2010}R. G. Unanyan, J. Otterbach, and M. Fleischhauer, Phys. Rev. Lett. {\bf 105}, 173603 (2010).
      \bibitem{Anderson1958}P.~W.~Anderson, 1958, Phys. Rev. {\bf 109}, 1492 (1958).
\bibitem{Kramer1993} B.~Kramer and A.~MacKinnon, Rep.\ Prog.\ Phys.\ \textbf{56}, 1469  (1993). 
     \bibitem{X_Zhang2007}C. Bai, X. Zhang, Phys. Rev. B {\bf 76}, 075430 (2007).
     \bibitem{Louie2008}C.-H. Park, L. Yang, Y.-W. Son, M. Cohen, S. G. Louie, Nat. Phys. {\bf 4}, 213-217 (2008).
     \bibitem{Pereira2008}M. Barbier, F. M. Peeters, P. Vasilopoulos, J. J. Milton Pereira, Phys. Rev. B {\bf 77}, 115446 (2008).
     \bibitem{Louie2008_2}C.-H. Park, Y.-W. Son, L. Yang, M. L. Cohen, S. G. Louie, Nano Lett. {\bf 8}, 2920 (2008).
     \bibitem{Lin2009}J. H. Ho, Y. H. Chiu, S. J. Tsai, M. F. Lin, Phys. Rev. B {\bf 79}, 115427 (2009).
     \bibitem{Fertig2009}L. Brey, H. A. Fertig, Phys. Rev. Lett. {\bf 103}, 046809 (2009).
     \bibitem{Peeters2010}M. Barbier, P. Vasilopoulos, F. M. Peeters, Phys. Rev. B {\bf 81}, 075438 (2010).
     \bibitem{SYZhu2010}L.-G. Wang and S.-Y. Zhu, Phys. Rev. B {\bf 81}, 205444 (2010).
     \bibitem{Louie2011}C.-H. Park, L. Z. Tan, and S. G. Louie, Physica E {\bf 43}, 651 (2011).
     \bibitem{Peeters2008}M. R. Masir, P. Vasilopoulos, A. Matulis, and F. M. Peeters, Phys. Rev. B {\bf 77}, 235443 (2008).
     \bibitem{Peeters2009}M. R. Masir, P. Vasilopoulos, and F. M. Peeters, Phys. Rev. B {\bf 79}, 035409 (2009).
     \bibitem{Sharma2009}S. Ghosh and M. Sharma, J. Phys.: Cond. Matt. {\bf 21}, 292204 (2009).
     \bibitem{Martino2009}L. Dell'Anna and A. D. Martino, Phys. Rev. B {\bf 79}, 045420 (2009).
     \bibitem{SJYang2008}Q.-S. Wu, S.-N. Zhang and S.-J. Yang, J. Phys.: Cond. Matt. {\bf 20}, 485210 (2008).
     \bibitem{Snyman2009}I. Snyman, Phys. Rev. B {\bf 80}, 054303 (2009).
     \bibitem{Louie2010}L. Z. Tan, C.-H. Park, and S. G. Louie, Phys. Rev. B {\bf 81}, 195426 (2010).
     \bibitem{Titov2010} S. Gattenl\"{o}hner, W. Belzig, and M. Titov, Phys. Rev. B {\bf 82}, 155417 (2010).
     \bibitem{Jonson2008}A. Isacsson, L. M. Jonsson, J. M. Kinaret, and M. Jonson, Phys. Rev. B {\bf 77}, 035423 (2008).
     \bibitem{Vozmediano2008}F. Guinea, M. I. Katsnelson, and M. A. H. Vozmediano, Phys. Rev. B {\bf 77}, 075422 (2008).
     \bibitem{Lichtenstein2008}T. O. Wehling, A. V. Balatsky, M. I. Katsnelson, and A. I. Lichtenstein, Europhys. Lett. {\bf 84}, 17003 (2008).
     \bibitem{Nov2004}K. S. Novoselov, A. K. Geim, S. V. Morozov, D. Jiang, Y. Zhang, S. V. Dubonos,
     I. V. Grigorieva, and A. A. Firsov, Science {\bf 306}, 666
     (2004).  
     \bibitem{Firsov2005}K. S. Novoselov, A. K. Geim, S. V. Morozov, D. Jiang, M. I. Katsnelson, I. V. Grigorieva, S. V. Dubonos, and A. A. Firsov,        Nature {\bf 438}, 197 (2005).
     \bibitem{Kim2005}Y. Zhang, Y. W. Tan, H. L. Stormer, and P. Kim, Nature {\bf 438}, 201 (2005).
     \bibitem{Zettl2008}J. C. Meyer, C. O. Girit, and M. F. Crommie, A. Zettl, Appl. Phys. Lett. {\bf 92}, 123110 (2008).
     \bibitem{Wintterlin2007}S. Marchini, S. G\"unther, and J. Wintterlin, Phys. Rev. B {\bf 76}, 075429 (2007).
     \bibitem{Miranda2008}A. L. Vazquez de Parga, F. Calleja, B. Borca, M. C. G. P. Jr, J. J. Hinarejo,
         F. Guinea, and R. Miranda, Phys. Rev. Lett. {\bf 100}, 056807 (2008).
     \bibitem{Sutter2008}P. W. Sutter, J.-I. Flege, and E. A. Sutter, Nature Mater. {\bf 7}, 406 (2008).
     \bibitem{Greber2008}D. Martoccia, P. R. Willmott, T. Brugger, M. Bj\"orck, S. G\"unther, C. M.
         Schlep\"utz, A. Cervellino, S. A. Pauli, B. D. Patterson, S. Marchini,
         J. Wintterlin, W. Moritz, and T. Greber, Phys. Rev. Lett. {\bf 101}, 126102 (2008).
     \bibitem{Michely2008}J. Coraux, A. T. N'Diaye, C. Busse, and T. Michely, Nano Lett. {\bf 8}, 565 (2008).
     \bibitem{Michely2008_2}A. T. N'Diaye1, J. Coraux, T. N. Plasa, C. Busse, and T. Michely, New J. Phys. {\bf 10}, 043033 (2008).
     \bibitem{Michely2009}I. Pletikosi\'c, M. Kralj, P. Pervan, R. Brako, J. Coraux, A. T. N'Diaye,
         C. Busse, and T. Michely, Phys. Rev. Lett. {\bf 102}, 056808 (2009).
     \bibitem{Avsar2011}A. Avsar \textit{et al}., Nano Lett. {\bf 11},
       2363 (2011).
     \bibitem{Egger2007}A. De Martino, L. Dell'Anna, and R. Egger, Phys. Rev. Lett. {\bf 98}, 066802 (2007).
     \bibitem{Neto2009}V. M. Pereira and A. H. Castro Neto, Phys. Rev. Lett. {\bf 103}, 046801 (2009).
     \bibitem{Peeters2009_3}M. R. Masir, P. Vasilopoulos, and F. M. Peeters, New J. Phys. {\bf 11}, 095009 (2009).
     \bibitem{ZDWang2009}S.-L. Zhu, D.-W. Zhang, and Z. D. Wang,
       Phys. Rev. Lett. {\bf 102}, 210403 (2009).
    \bibitem{Nori2009}Y.~P. Bliokh, V. Freilikher, S. Savel'ev, and F. Nori, Phys. Rev. B {\bf 79}, 075123 (2009).
     \bibitem{Soukoulis2008}P. Marko\v s and C. M. Soukoulis, {\it Wave Propagation: from electrons to photonic crystals and left-handed materials} (Princeton University Press, Princeton and Oxford, 2008), Chap. 1-4.
     \bibitem{Delande2010} C.~A.~M\"uller and D.~Delande, chap.~9 in
       C.~Miniatura et al. (eds), \textit{Les Houches 2009 - Session XCI: Ultracold Gases and Quantum Information}
(Oxford University Press, Oxford 2011);
       arXiv:1005.0915
\bibitem{GalindoPascualQM1} A.~Galindo, P.~Pascual, \textit{Quantum Mechanics I}
  (Springer, Berlin, 1990).
     \bibitem{Furstenberg}H. Furstenberg and H. Kesten, Ann. Math. Statist. {\bf 31}, 457 (1960);
          H. Furstenberg, Trans. Amer. Math. Soc. {\bf 108}, 377
          (1963).
     \bibitem{Abrikosov1981} A.~A.~Abrikosov, 
Sol.~State Comm.~\textbf{37}, 997 (1981).
\bibitem{Berry1997} M.~V.~Berry and S.~Klein, 
Eur. J. Phys. \textbf{18}, 222 (1997).
     \bibitem{Derrida1987}B. Derrida, K. Mecheri, and J. Pichard,
       J. Phys. France {\bf 48}, 733 (1987).

\bibitem{Derrida1984} B.~Derrida, E.~Gardner, and J.~Physique \textbf{45}
  1283 (1984).

\bibitem{Izrailev1998} F.~M.~Izrailev, S.~Ruffo, and L.~Tessieri,, J.\
  Phys.\ A: Math.\ Gen. \textbf{31}, 5263 (1998).
\bibitem{Peeters2009_2} M. Barbier, P. Vasilopoulos, and
        F. M. Peeters, Phys. Rev. B {\bf 80}, 205415 (2009).

\bibitem{Izrailev2001}
F.~M.~Izrailev, A.~A.~Krokhin, S.~E.~Ulloa,
Phys. Rev. B \textbf{63},  041102 (2001).

\bibitem{Krokhin2002} A.~Krokhin, and F.~M.~Izrailev, U.~Kuhl,
  H.-J.~St\"ockmann, S.~E.~Ulloa, Physica E \textbf{13}, 695  (2002).

\bibitem{Lugan2009}
P.~Lugan, A.~Aspect, L.~Sanchez-Palencia, D.~Delande, B.~Gr\'emaud, C.~A.~M\"uller, C.~Miniatura,
Phys.\ Rev.\ A \textbf{80}, 023605 (2009).


  \bibitem{Aihua2010}A. Zhang, Z. Dai, L. Shi, Y. P. Feng, and C. Zhang, J. Chem. Phys {\bf 133}, 224705 (2010).


\end{thebibliography}
\end{document}